\documentclass[doublecol]{epl2}
\usepackage{graphicx}

\title{
Future Directions for Active Matter On Ordered Substrates 
} 

\author{
C. Reichhardt$^1$, A. Lib{\' a}l$^2$, and C. J. O. Reichhardt$^1$ 
} 

\institute{
$^1$Theoretical Division and Center for Nonlinear Studies,
Los Alamos National Laboratory, Los Alamos, New Mexico 87545, USA\\
$^2$Mathematics and Computer Science Department, Babe{\c s}-Bolyai University,
Cluj-Napoca 400084, Romania}

\abstract{
Active matter is a term encompassing particle-based assemblies with some form of self-propulsion, including certain biological systems as well as synthetic systems such as artificial colloidal swimmers, all of which can exhibit a remarkable variety of new kinds of nonequilibrium phenomena. A wealth of non-active condensed matter systems can be described in terms of a collection of particles coupled to periodic substrates, leading to the emergence of commensurate-incommensurate effects, Mott phases, tribology effects, and pattern formation. It is natural to ask how such phases are modified when the system is active. Here we provide an overview and future directions for studying individual and collectively interacting active matter particles coupled to periodic substrates, where new types of commensuration effects, directional locking, and active phases can occur. Further directions for exploration include locking effects, the realization of active solitons or active defects in incommensurate phases, active Mott phases, active artificial spin ice, active doping transitions, active floating phases, active surface physics, active matter time crystals, and active tribology.  
}
\begin{document}

\maketitle

\section{Introduction}
Active matter systems, composed of particles or components that 
have some form of self-propulsion
\cite{Marchetti13,Bechinger16}, are relevant to a wide 
variety of areas
including biological systems \cite{Marchetti13,Bechinger16}, 
synthetic systems
such as artificial colloidal swimmers
undergoing chemical catalysis
\cite{Marchetti13,Bechinger16},
crowds \cite{Helbing00}, and robotic systems \cite{Wang21}.
There are also continuum versions of active 
matter such as cell motion and active nematics
\cite{Marchetti13,Giomi13a,DeCamp15}.
In the case of particle based active matter systems,
the individual particles show anomalous diffusion
due to the finite persistence or run length of their motion.
For 
collectively interacting
active particle systems, other types of behaviors arise,
with one of the best known examples being motility induced phase 
separation where the particles form a high density solid phase coexisting 
with a low density gas when the activity is large enough
\cite{Fily12,Redner13,Palacci13,Buttinoni13,Reichhardt15,Cates15}.
When active particles are placed in confinement,
they can accumulate
along walls or boundaries \cite{Tailleur09,Fily15,Sepulveda17},
and the resulting active pressure forces can show marked differences 
from forces observed in equilibrium systems
\cite{Solon15,Ray14,Ni15,Kjeldbjerg21}.
With asymmetric boundaries,
active ratchet effects can arise \cite{Galajda08,Reichhardt17a}. 
There have also been a variety of studies of active matter systems
in disordered media where the disorder can break up
the active matter clusters or lead to anomalous diffusion
\cite{Reichhardt14,Chepizhko15,Morin17,Bhattacharjee19a}. 

A class of nonactive systems that has been
extensively studied on the individual and
collectively interacting particle level
involves the coupling of particles to a periodic substrate 
\cite{Reichhardt17}, which may take the form of a
periodic array of obstacles or trapping sites
created optically or with nanostructures.
A series of commensuration effects 
occur when the number of particles or the spacing between
particles matches the substrate periodicity.
At commensuration, the system can be highly ordered
and the motility is
strongly reduced.
Such effects have been observed
for ions in periodic substrates \cite{Benassi11}, 
cold atoms in optical arrays \cite{Buchler03,Lewenstein07,Gross17},
colloidal particles \cite{Brunner02,Brazda18},
surface physics \cite{Coppersmith82,Bak82}, 
and superconducting vortex systems \cite{Harada96,Reichhardt98a}.
Other effects including Mott transitions, novel crystalline ordering,
and melting
have also been studied as a function of the substrate strength.
In more complex periodic substrate geometries,
frustration effects can arise even under commensurate
conditions, leading to the formation
of excitations such as monopoles of the type found
in two-dimensional artificial spin ice systems
\cite{OrtizAmbriz19}. 
Under an applied drive, particles on periodic substrates exhibit numerous 
types of depinning, sliding, and frictional behavior,
and in incommensurate states, depinning transitions can occur via
the nucleation and motion of solitons
\cite{Bohlein12,Vanossi13,Reichhardt17}. 

Here we give a perspective on active matter systems interacting with
periodic or ordered substrates, highlighting what has been
done so far and 
emphasizing the new directions to pursue in this area.
There could be active matter Mott phases at commensurate states, or
the activity could enhance localization.
At incommensurate states, there could be motion
or  depinning of solitons due to the activity, leading to the emergence of
new types of active solitons.
In chiral active 
matter systems, it could be possible to realize active 
versions of spin systems and spin ordered states.
It may also be possible to create dynamical states that are periodic in time 
in order to realize an active matter version of time crystals.
Other directions include studying the
impact of activity on tribology effects such as friction, lubrication,
and wear by exploring the
sliding of active matter over periodic sites or
by enclosing active matter particles between sliding plates.
It would also be interesting to add activity to
well known models for particles on periodic
substrates in both
classical and quantum contexts to 
examine active matter versions of spin ice, active Mott phases,
active insulator to fluid transitions, active doping, and so forth.
Other directions include studying active continuum models such as
active charged systems, active nematics and active polymers.

\section{Modeling Different Types of Substrates}

\begin{figure}
\onefigure[width=\columnwidth]{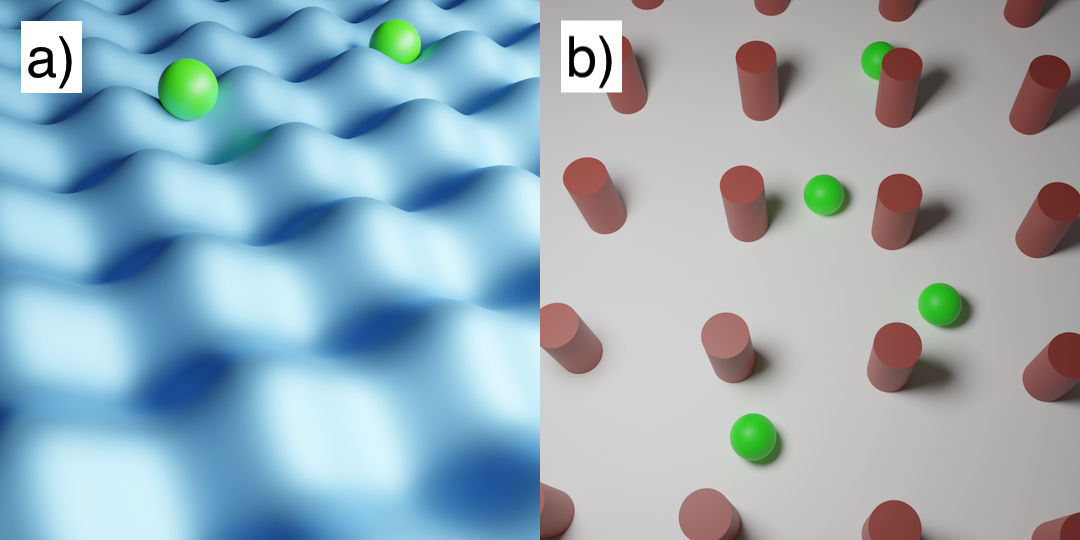}
\caption{
Illustration of two types of substrates.    
(a) An egg-carton substrate for active particles, where there are well-defined
trapping locations that 
can capture one or more particles (green).
(b) An array of obstacles or posts (red) where active particles (green)
can move freely between obstacles but collide with or are guided by the obstacles. 
These different substrate types produce different effects. 
}
\label{fig:1}
\end{figure}

The simplest representation of particle-based active matter employs
self motile disks 
with a finite short range repulsion and a motor force;
however, other more complicated interactions can also be included such as
elongated particles, hydrodynamic effects, and even charged particles.
Typically, active matter particle systems are modeled as overdamped systems with
a motor force for propulsion, interaction forces between the particles, and
a substrate interaction:
\begin{equation} 
	\eta \frac{d{\bf r}_{i}}{dt} = {\bf F}_{\rm inter} + {\bf F}_{\rm sub} + {\bf F}_{m}.
\end{equation}
The particle position is ${\bf r}_{i}$ and its velocity is 
${\bf v}_{i} = {d {\bf r}_{i}}/{dt}$, while the damping constant $\eta$ in the examples
shown here is set to $\eta=1$.
The particle-particle interaction force for the case of $N_a$ hard disk particles is given by
${\bf F}_{\rm inter}=\sum_{j\neq i}^{N_a}k(|{\bf r}_{ij}|-2r_a)
\Theta(|{\bf r}_{ij}|-2r_a){\bf \hat{r}}_{ij}$
where $r_a$ is the particle radius, $\Theta$ is the Heaviside step function,
${\bf r}_{ij}={\bf r}_i-{\bf r}_j$, and
${\bf \hat{r}}_{ij}={\bf r}_{ij}/|{\bf r}_{ij}|$.
The substrate interaction is ${\bf F}_{\rm sub}$ and can be of
different forms as described below.
The motor force ${\bf F}_m$ represents either
run-and-tumble motion or driven diffusion.
In either case there is a run length or persistence length,
called the activity length $l_a$,
which determines how far the particle moves in one direction in the absence of collisions. 
If there are multiple particles, an additional length scale appears in the form of
the average distance $a$ between the particles.
In two-dimensional systems, there are two general types of attractive interaction
potential. The egg-carton potential illustrated in Fig.~\ref{fig:1}(a)
is composed of
a periodic array with lattice constant $a_s$
of traps that can capture one or more particles.
Alternatively, an array of obstacles can be used as shown in Fig.~\ref{fig:1}(b).
Other types of substrates are possible,
including the muffin tin substrate in which the well defined pinning sites
have flat spaces between them, quasi-one-dimensional substrates in the form
of aligned troughs, and quasiperiodic substrates.
For the substrate in Fig.~\ref{fig:1}(a), the minima
act as trapping sites with a maximum trapping force of $F_{p}$
such that a single particle becomes trapped in a site when $F_{m}<F_p$.
Even when $F_{m} > F_{p}$, if the persistence length $l_a$ is very small,
a particle can remain trapped by a site for long periods of time.
In principle, when $l_{a} > a_{s}$ and $F_{m} > F_{p}$, a particle should be continuously 
hopping from one site to the next.
In contrast, for the obstacle substrate shown in Fig.~\ref{fig:1}(b),
single particles are
generally not trapped but move through
the flat space between obstacles.
Encounters with the
obstacles can impede or guide the particle motion.
This shows that different types of substrate produce distinct effects. 

\begin{figure}
  \begin{minipage}{\columnwidth}
    \onefigure[width=\columnwidth]{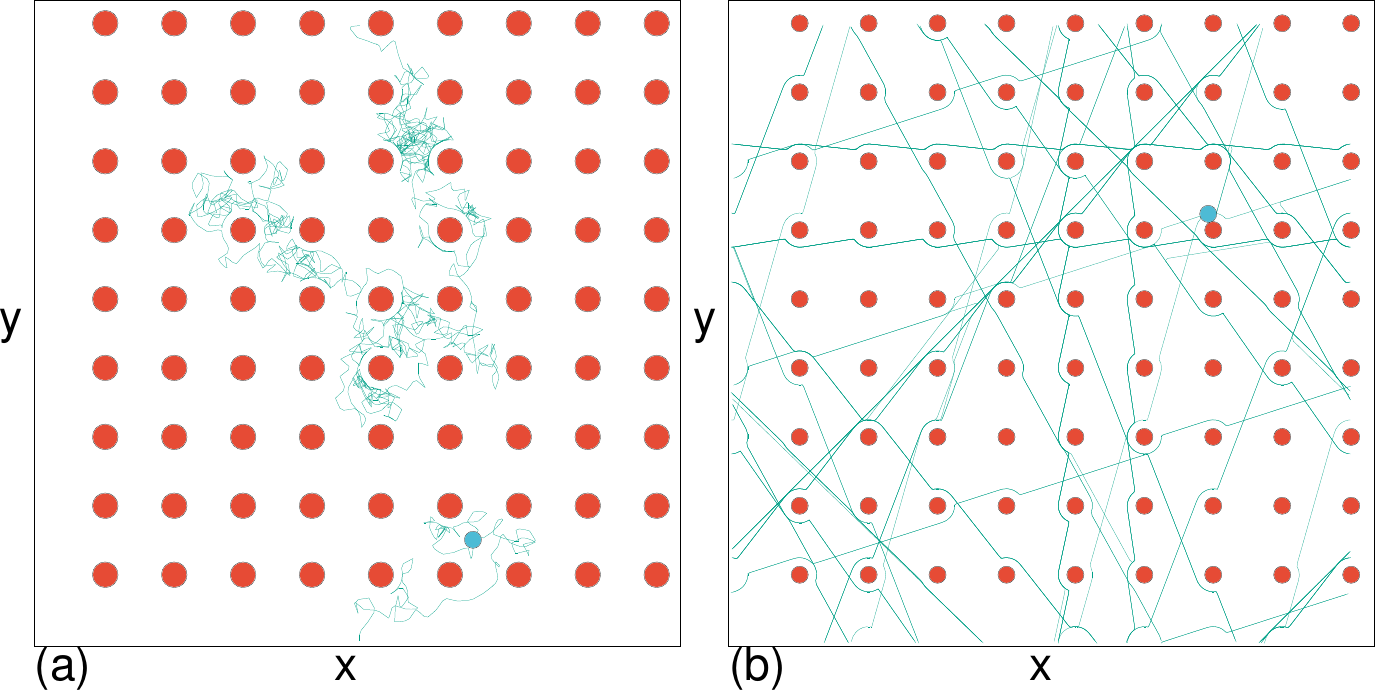}
  \end{minipage}\hfill
  \begin{minipage}{\columnwidth}
    \onefigure[width=\columnwidth]{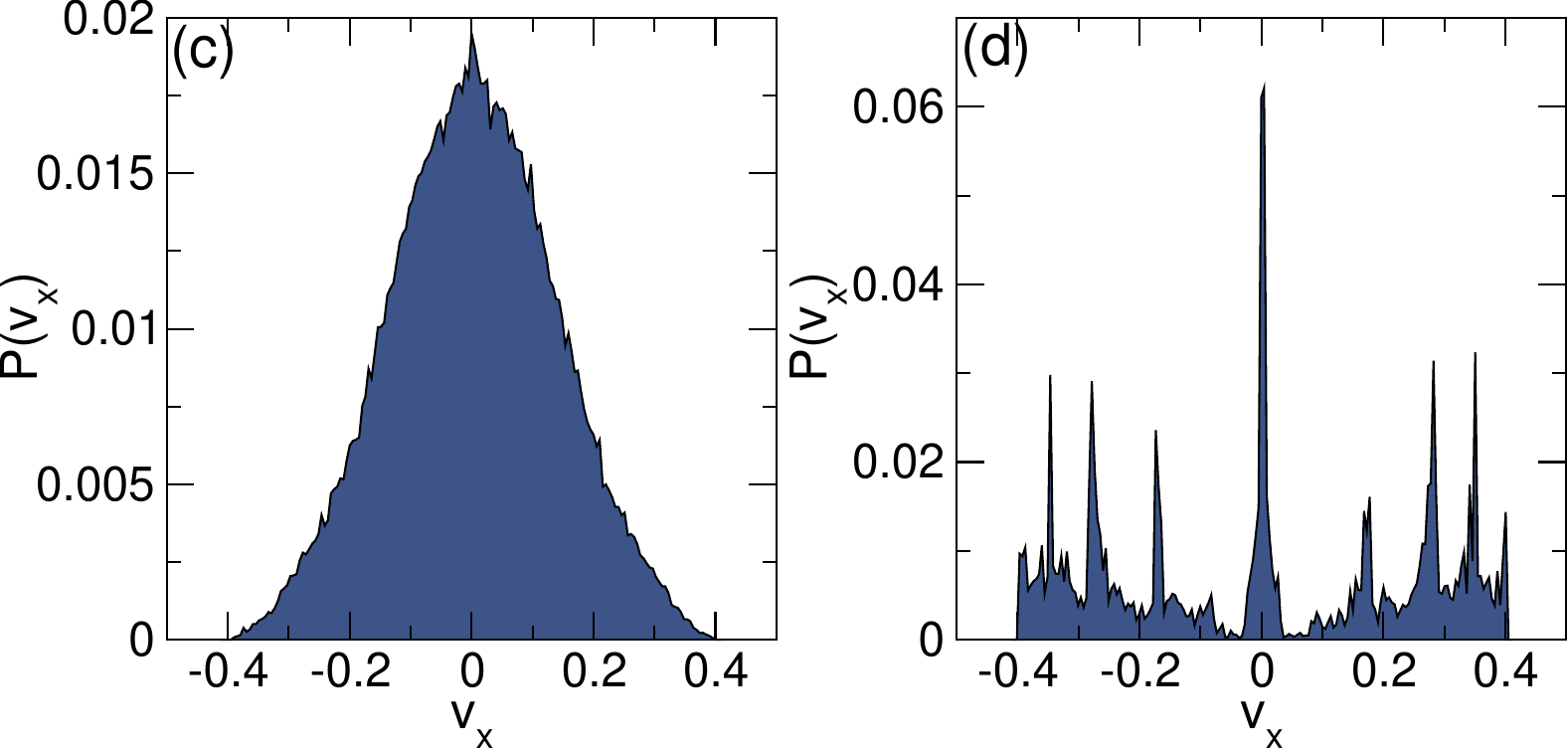}
  \end{minipage}\hfill
\caption{
(a) An array of posts (red) and trajectories (green) of  a
single active particle in the Brownian limit.
(b) The same but for a system of run-and-tumble particles with a run length
$l_a=20a_s$, where $a_s$ is the lattice constant of the obstacle array.
The motion locks to specific symmetry directions
of the substrate.
(c) The distribution $P(v_x)$ of
instantaneous $x$-direction velocities from the system in panel (a) 
has Gaussian features.
(d) $P(v_x)$ from the system in panel (b) shows several peaks that correspond
to the locking of the particle motion to certain
substrate symmetry directions.
Results are from Ref.~\cite{Reichhardt20}.
Reprinted with permission from C. Reichhardt {\it et al.}, Phys. Rev. E {\bf 102},
042616 (2020). Copyright 2020 by the American Physical Society.
}
\label{fig:2}
\end{figure}

Even in a system containing only a single active particle,
the presence of a periodic substrate alters
the motion of the particle.
In Fig.~\ref{fig:2}(a,b) we illustrate an array of posts along with
the trajectory of a single active disk for the system studied in
Ref.~\cite{Reichhardt20}. 
When the run length $l_a$ is very short, as in Fig.~\ref{fig:2}(a),
the system behaves in a Brownian manner and the particle
trajectory
fills space uniformly over time.
In Fig.~\ref{fig:2}(b),
the run length $l_a=20a_s$, where
$a_s$ is the spacing between obstacles.
The motion is constrained to follow symmetry directions of the
lattice at angles $\theta=0^\circ$, $45^\circ$, and $90^\circ$ to the $x$ axis.
The distribution $P(v_x)$ of instantaneous $x$-direction velocities for the small run
length particle from Fig.~\ref{fig:2}(a) appears in
Fig.~\ref{fig:2}(c)
and has a Gaussian form, while in Fig.~\ref{fig:2}(d),
$P(v_x)$ for the $l_a=20a_s$ particle from Fig.~\ref{fig:2}(b)
has
a series of spikes
corresponding to the different locking directions.

For a system with an egg carton potential, different kinds of
single particle dynamical effects could arise for $F_m/F_p>1.0$
whenever there is
matching between the persistence length
and the substrate lattice spacing such that $l_{a}/a_{s} = n$ with $n$ integer,
leading to an oscillating mobility as a function of varying $F_m$.

\section{Commensuration effects}

In Fig.~\ref{fig:1}(a), it is clear that commensuration effects appear when
there is a matching density of one particle per substrate trap.
It would be interesting to explore the
effective mobility of the system as a function of the filling ratio
$f$ of the number of active particles $N_{a}$ to  the number of
substrate minima  $N_{s}$,  $f = N_{a}/N_{s}$.
These effects will also depend on the
values of the motor force and the persistence length. 
When $F_{m}/F_p > 1.0$, in the single particle limit the
particle readily hops from one trap to another;
however, for $N_{a}/N_{s} = 1.0$, all of the wells are occupied and there
is no free trap available to accommodate hopping without
particle-particle collisions,
causing a strong reduction of the mobility and producing an
effectively insulating state.
One question is whether there is an optimal activity that would enhance the
insulating behavior.
Another possibility is to dope the $f=1$ filling with interstitial active
particles or holes, or to replace some of the active particles by
non-active particles.
For example, it is possible that the addition of non-active particles 
would increase the net mobility of the system since the
non-active particles may have difficulty finding and being trapped by the
pinning sites, leaving holes available where active particles could undergo
hopping transport.
It is also interesting to consider
whether different kinds of ordered states would arise at higher filings of
$f = 2, 3...n$
and whether motility induced phase separation
could still occur at these higher fillings or if it would be suppressed by the substrate.  

\begin{figure}
  \begin{minipage}{\columnwidth}
    \onefigure[width=\columnwidth]{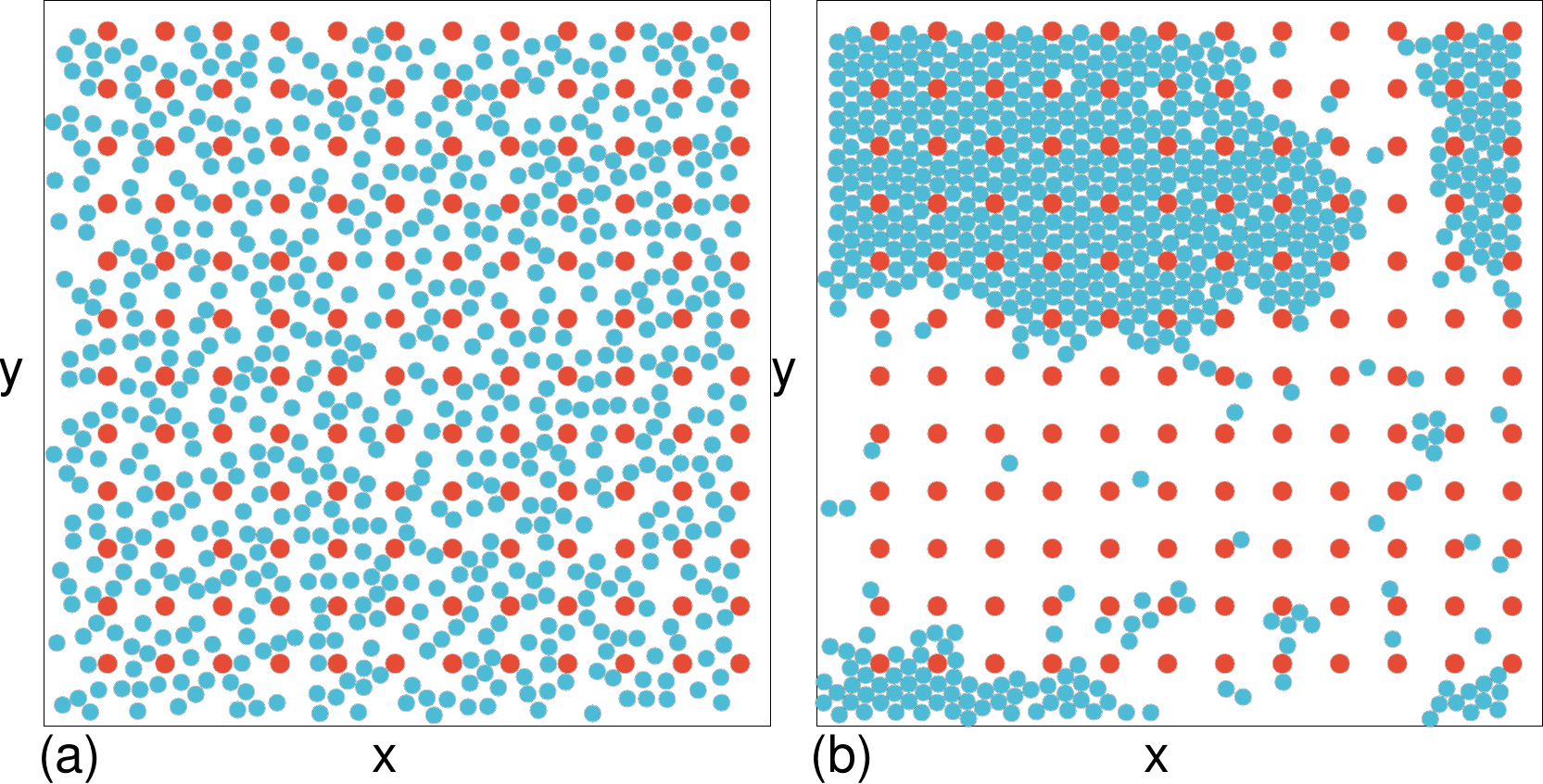}
  \end{minipage}\hfill
  \begin{minipage}{\columnwidth}
    \onefigure[width=0.8\columnwidth]{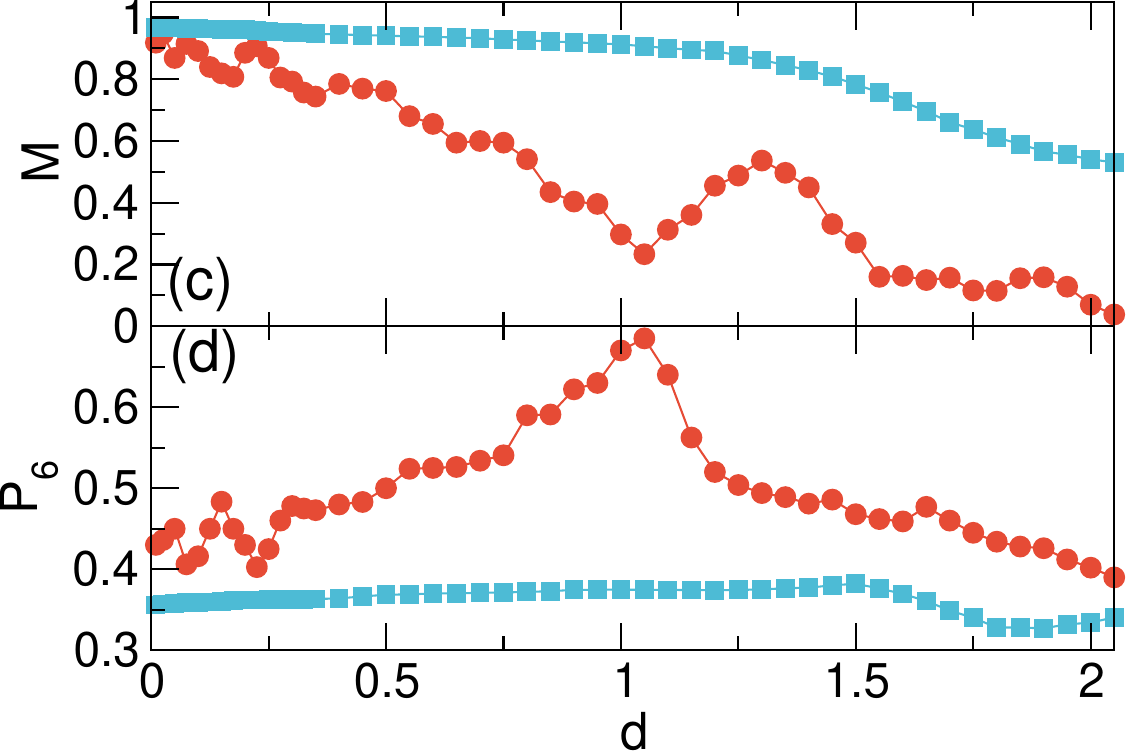}
  \end{minipage}\hfill
\caption{
(a)  Mobile disks (blue) interacting with a
square array of obstacles (red) for a system
in the Brownian limit of small $l_a$ for an active disk density of $\phi = 0.32$,
where the system forms a uniform fluid.
(b) The same system at a long run length $l_a$ where 
local commensurate clusters with triangular ordering appear. 
(c) Red: The mobility of the system versus $d$ showing a dip
near $d=1.0$. The blue curve is for a sample with no obstacle array.
(d) Red: The fraction $P_6$ of particles with six neighbors vs obstacle diameter
$d$ for the system in panel (b). A peak appears
near $d = 1.0$ at the same value where $M$ has a dip; this
corresponds to the ordered state shown
in panel (b).
The blue curve is for a sample with no obstacle array.
Results are from Ref.~\cite{Reichhardt21}.
Reprinted with permission from C. Reichhardt {\it et al.}, Phys. Rev. E {\bf 103},
022602 (2021). Copyright 2021 by the American Physical Society.
}
\label{fig:3}
\end{figure}

There can also be commensurability effects
for the periodic obstacle array.
If the active particles have radius $r_a$, commensuration effects
can occur when the spacing between the obstacles is
an integer multiple of the active particle radius $2r_{a}$,
and there can be other types of geometric arrangements that
allow for the disks to adopt a periodic spacing between the obstacles.
In a nonactive system,
commensuration effects occur when $N_{a}/N_{s} = n$
with $n$ integer \cite{Reichhardt17,Harada96,Reichhardt98a,Bak82}, 
implying that the system must have a uniform density.
Active matter can exhibit motility-induced phase separation in which a
region of high density coexists with a low density gas of particles,
so commensuration effects can occur even when
$N_{a}/N_{s}<1$.
This phenomenon 
has already been studied for active run-and-tumble particles interacting with
a square array of obstacles \cite{Reichhardt21}.
Figure~\ref{fig:3}(a) shows a system of mobile disks 
in a square obstacle array in the Brownian limit of small $l_a$.
In this case, the area covered by the disks is $\phi = \pi r_a^2/L^2 = 0.32$,
where $L$ is the length of one side of the sample.
This is well below the jamming density of $\phi=0.9$, and
as the figure illustrates, the system forms
a uniform fluid.
In Fig.~\ref{fig:3}(b), the same system with a
long running length or large $l_a$
forms a motility induced phase separated state in the absence of a substrate. In
the presence of the substrate,
local areas of high density
form a commensurate triangular ordering that coexists with 
a low density gas.
The triangular ordering occurs when
the obstacle diameter is chosen such that an integer number of disks can 
fit between the obstacles.
For other obstacle diameters,
a cluster can still form when the activity is large
but it is
a lower density disordered cluster.
The active commensurate states
can be detected
by
measuring the ordering as a function of
obstacle diameter $d$, as shown in
Fig.~\ref{fig:3}(d) where the fraction of sixfold coordinated particles $P_6$ is plotted
versus $d$.
There is a peak at 
$d = 1.0$ where the crystalline state shown in Fig.~\ref{fig:3}(b) occurs.
For smaller $d$, the obstacles break up the clusters.
Figure~\ref{fig:3}(c) shows the corresponding mobility
$M=N_a^{-1}\sum_{i}^{N_a}v_x$ versus 
$d$ for the same system.
The peak in $P_{6}$ corresponds to the lowest mobility, 
indicating that the commensurate cluster is pinned.
There are also a series of other bumps and dips
corresponding to different kinds of commensurate-incommensurate ordering.
The peak in the mobility near $d = 1.35$
appears when the system forms a loosely packed disordered 
solid.
There are many future directions to study. For example,
one could look at mixtures of active and passive particles  to see 
if they would mix or dimix.
Also of interest would be a mixture 
of elongated active particles such as bacteria or anisotropic colloids 
where one length scale matches the substrate
lattice constant and the other does not.
Such a system could form a commensurate cluster state for only one of the
two types of particles, producing
a much richer variety of ordering. 

It would also be interesting to explore more complex periodic substrates.
For example,
there are geometries known as artificial spin ice that can be
modeled as particles coupled to a specific type of
periodic substrate that gives rise to a frustrated ground state and supports excitations
such as monopoles \cite{OrtizAmbriz19}.
In active systems, the monopoles could act as active emergent particles, 
and active order to disorder transitions could be studied as a function of activity. 

\begin{figure}
\onefigure[width=\columnwidth]{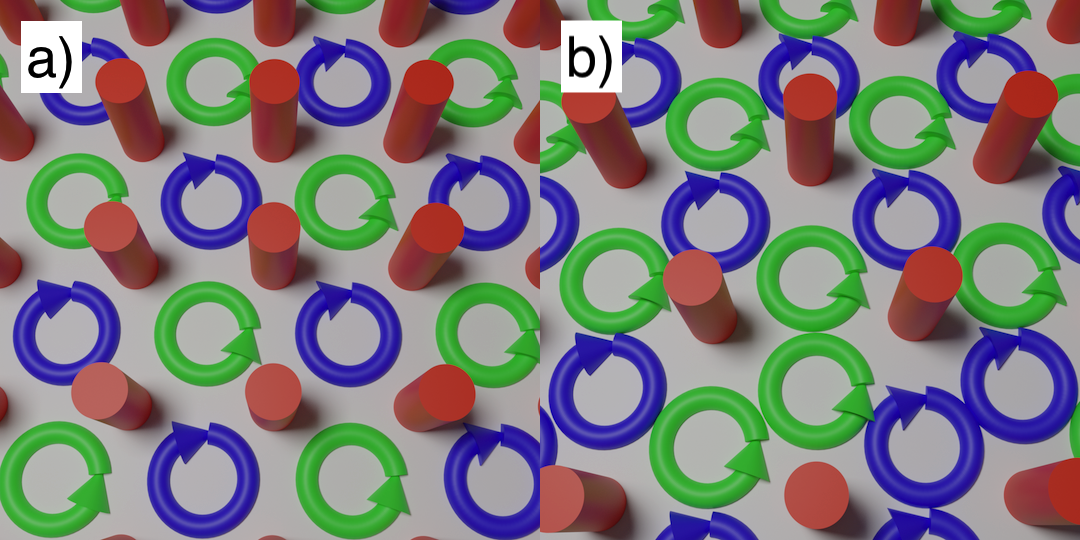}
  \caption{ (a) A square array of obstacles (red)
interacting with a binary assembly of chiral swimmers
(blue: clockwise; green: counterclockwise)    
at 1:1 matching where the swimmers can form a
checkerboard ordered state.
(b) The same but for a triangular substrate array where
the system forms a disordered state. 
}
\label{fig:4}
\end{figure}

Another form of active matter
is chiral swimmers or particles that move in circular paths
\cite{Lowen16,Han17,Reichhardt19}.
The direction
of the chirality can be viewed
as an effective spin degree of freedom for 
active particles,
allowing spinners on a periodic substrate to be used as active
matter versions of various spin models.
For example, in a system of 
clockwise and counterclockwise swimmers
placed on a square periodic lattice,
if long range interactions couple 
different plaquettes of the sample,
an ordered checkerboard state could form as shown in Fig.~\ref{fig:4}(a).
On the other hand, if the
swimmers are placed on a triangular lattice,
they will generally be frustrated
and disordered as shown in Fig.~\ref{fig:4}(b), although 
other orderings could  arise such as stripe phases.
Another interesting possibility is that 
chiral or nonchiral active matter on a periodic substrate could synchronize to form
dynamical states that repeat as a function of time, producing
an active matter version of classical
time crystals \cite{Shapere12,Yao20,Libal20}.
Other effects to explore include phase locking, chaos, or
intermittent states similar to those found for coupled oscillators. 

\begin{figure}
    \onefigure[width=\columnwidth]{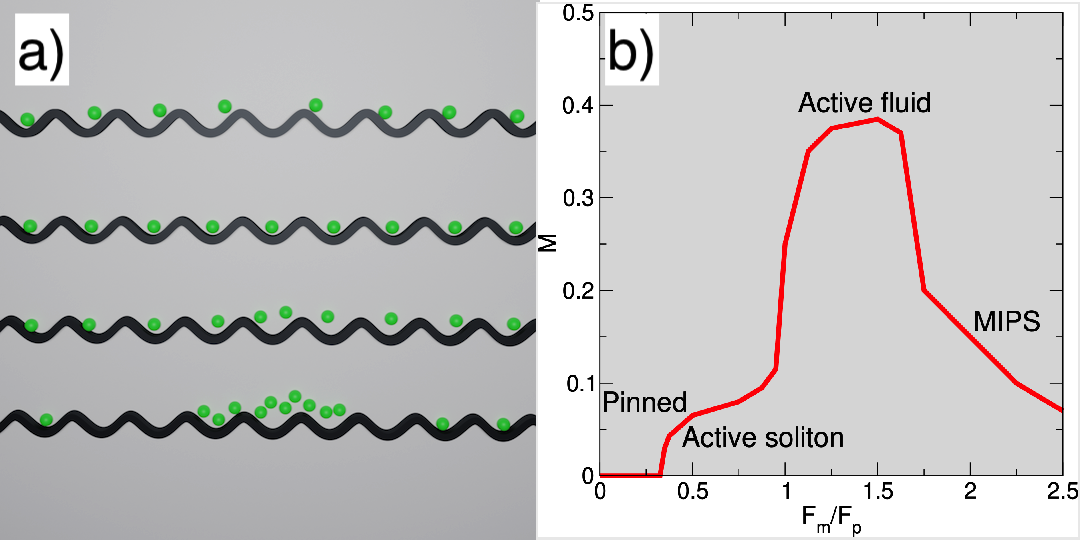}
\caption{ 
(a) Schematics of an active matter system on a periodic substrate where
there is a kink or soliton present. The soliton could show active motion
while the other active particles remain pinned.
Top: a vacancy soliton. Upper middle: a commensurate state. Lower middle:
an interstitial soliton. Bottom: a motility induced phase separated (MIPS) state.
(b) Schematic showing possible behavior
of the mobility $M$ of an incommensurate active matter system
as a function of increasing  motor force $F_M$ showing a pinned phase,
an active soliton phase, an active fluid, and MIPS
states. 
}
\label{fig:5}
\end{figure}

\section{Incommensurate phases} 
In systems with periodic substrates, a rich variety of effects appear
when the number of particles is not an integer multiple of
the number of substrate minima.
This can lead to the emergence of a disordered or fluid state.
Just  away from commensuration, these systems
typically form kink states where there
are either additional particles or missing particles that 
behave like localized excitations. Under a drive,
these kinks have a much lower depinning threshold
than the commensurate particles \cite{Reichhardt17,Bohlein12,Vanossi13}.  
For active matter systems, it is not known whether
such kinks can form and
whether they would have active dynamics themselves.
In Fig.~\ref{fig:5}(a) we show schematics of a quasi-one-dimensional system to
demonstrate
how kinks or antikinks
could form on a periodic substrate that
is just off commensurability.
If the motor force of the active particles
is less than the trapping force for an isolated particle,
the kinks experience a lower barrier at the edge of the pinning site
and could effectively depin and move
through the background of pinned particles. 
Unlike kinks in a nonactive system that generally 
move only in the direction of an
externally applied force,
active kinks or active solitons could move in any direction.
The 
first question is whether the kinks would more frequently move along
symmetry directions of the substrate, similar to what is observed for
symmetry locking in driven states.
Additionally,  if two kinks collide, would they behave like solitons
and pass though one another, or would they scatter 
into other directions?
If kinks are like active particles themselves,
would they be able to undergo motility induced phase separation?
In non-active systems, simulations and experiments
on kink depinning as a function of increasing drive
show  a multiple step depinning
process \cite{Reichhardt17,Bohlein12,Vanossi13}.
For active systems, it should be possible to
increase the motor force and observe different motility 
regimes, as illustrated schematically in Fig.~\ref{fig:5}(b).
The first regime would contain only moving kinks, while at higher forces
all of the active particles would become mobile.
The mobility may drop again for higher motor forces
if the system can enter a motility induced phase separated state
of the type shown on the bottom of Fig.~\ref{fig:5}(a).

\begin{figure}
    \onefigure[width=\columnwidth]{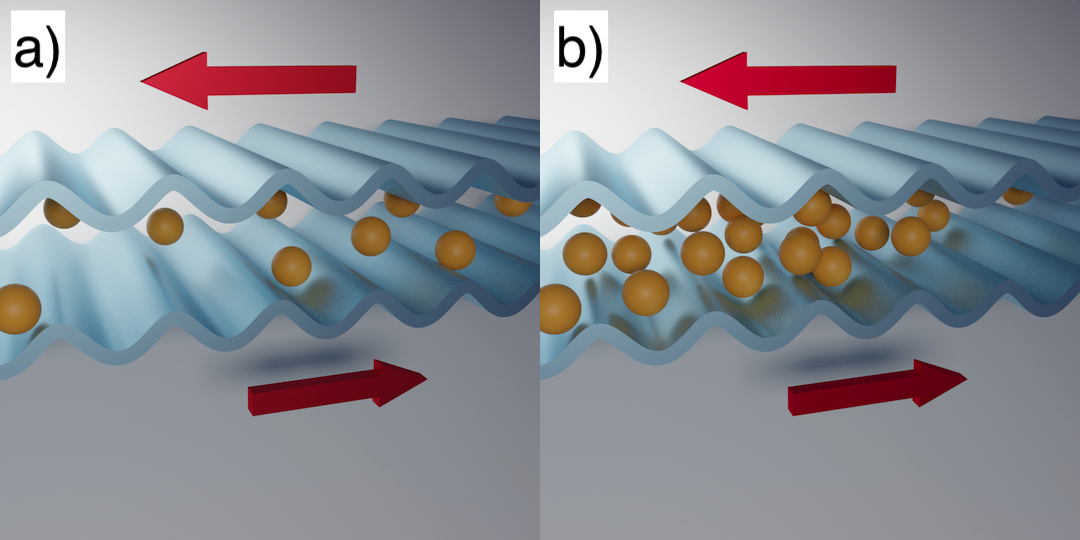}
        \caption{ An example of active matter tribology showing two periodic  plates
	  sliding past one another with active particles (yellow)
          in the space between the plates. Arrows indicate the direction of motion of
          each plate.
          (a) Low density of active particles. (b) Higher density of active particles.
	}
\label{fig:6}
\end{figure}

Tribology effects \cite{Bhushan13} could be studied
with active matter by using two periodic surfaces, one active and
one passive, that are driven
past each other.
Another possibility would be to consider the possible lubricating effects
of adding active matter to the space between two 
sliding planes, as shown schematically in Fig.~\ref{fig:6}.
In a nonactive system,
the friction can often be modified depending on  the density and
type of particles placed between the planes;
here, one could explore what would happen if the particles are active and
whether the friction could be controlled by changing the activity. 

\section{Discussion}
So far we have only described active matter composed of hard particles.
Additional
effects could arise for soft or deformable active particles or
for continuum active nematic phases, which
contain large scale defects that have a well defined spacing
\cite{Giomi13a,DeCamp15}.
It wold be interesting to see whether these defects could be
locked by an array of obstacles if the obstacles have a similar 
spacing, or if an incommensurate spacing could induce or suppress the
formation of such defects.
Other possible systems include active polymers with a length that could match
the periodicity of the substrate.
Since many of the systems that have previously been studied involve
charged particles or particle-particle interactions that are longer range
than hard spheres,
a natural future direction is to study
active charged systems or
active particles with Yukawa interactions.
It could also be possible to make particles with a Bessel function
interaction
active.
There are some efforts underway currently
to create such systems experimentally, 
allowing more direct comparisons to
the commensurate states and dynamics found
in colloidal and superconducting vortex systems
without activity. 
Another approach would be simply to take many of the known
models for systems that have been studied
on periodic lattices and add an activity term.
This could be done for both classical and even quantum
systems, where correlated noise or a persistence length could be added.
There may be a
way to create such systems experimentally using active optical feedback traps.

An additional area is to consider systems with a combination of periodic
pinning and an asymmetry in order to create active ratchet effects that can
be impacted by commensuration. Some studies along these lines have been
performed in nonactive systems \cite{Derenyi95,deSouzaSilva06,Lu07}
as well as for active matter on periodic asymmetric substrates
\cite{Potiguar14,Lozano16,McDermott16,Reichhardt17a}.
Other directions are to study chiral active matter on periodic asymmetric
substrates \cite{Reichhardt13a,Reichhardt20b}
where Hall effects and odd viscosity can arise.
Finally, the flocking of swarming models of self-propelled particles
could be examined in the presence of a periodic
substrate \cite{Vicsek95,Drocco12}.

\section{Summary}
Active systems have opened a new field of physics and materials
and represent a new state of matter.
In nonactive systems there is an extensive amount of 
phenomena that has been observed under coupling
to a periodic substrate, such as as commensuration effects,  formation
of kinks and anti-kinks, diffusion, Mott and fluid like phases,
tribology, and pattern formation
for both hard and soft matter systems.
Examining active matter on periodic substrates
is a natural direction for further work
where new types of commensurate effects, directional locking, ordering
and frustrated order could be observed.
We highlighted several of the directions that could be studied for
different types of periodic substrates, including
commensuration effects for increasing activity and motor forces,
dynamical commensuration effects when the persistence length is
a multiple of the substrate periodicity,
active solitons or active excitations in incommensurate phases,
new ways to control friction,
and new types of pattern formation.
It would also  be interesting to incorporate activity into 
already existing models in both classical and
quantum systems coupled to periodic substrates.

\acknowledgments
We gratefully acknowledge the support of the U.S. Department of
Energy through the LANL/LDRD program for this work.
This work was supported by the US Department of Energy through
the Los Alamos National Laboratory.  Los Alamos National Laboratory is
operated by Triad National Security, LLC, for the National Nuclear Security
Administration of the U. S. Department of Energy (Contract No. 892333218NCA000001).
AL was supported by a grant of the Romanian Ministry of Education
and Research, CNCS - UEFISCDI, project number
PN-III-P4-ID-PCE-2020-1301, within PNCDI III.

\bibliographystyle{eplbib}
\bibliography{mybib}

\begin{thebibliography}{10}
\expandafter\ifx\csname url\endcsname\relax\def\url#1{\texttt{#1}}\fi

\bibitem{Marchetti13}
\Name{Marchetti M.~C., Joanny J.~F., Ramaswamy S., Liverpool T.~B., Prost J.,
  Rao M. \and Simha R.~A.} \REVIEW{Rev. Mod. Phys.}{85}{2013}{1143}.

\bibitem{Bechinger16}
\Name{Bechinger C., Di~Leonardo R., L\"owen H., Reichhardt C., Volpe G. \and
  Volpe G.} \REVIEW{Rev. Mod. Phys.}{88}{2016}{045006}.

\bibitem{Helbing00}
\Name{Helbing D., Farkas I. \and Vicsek T.} \REVIEW{Nature
  (London)}{407}{2000}{487}.

\bibitem{Wang21}
\Name{Wang G., Phan T.~V., Li S., Wombacher M., Qu J., Peng Y., Chen G.,
  Goldman D.~I., Levin S.~A., Austin R.~H. \and Liu L.} \REVIEW{Phys. Rev.
  Lett.}{126}{2021}{108002}.

\bibitem{Giomi13a}
\Name{Giomi L., Bowick M.~J., Ma X. \and Marchetti M.~C.} \REVIEW{Phys. Rev.
  Lett.}{110}{2013}{228101}.

\bibitem{DeCamp15}
\Name{DeCamp S.~J., Redner G.~S., Baskaran A., Hagan M.~F. \and Dogic Z.}
  \REVIEW{Nature Mater.}{14}{2015}{1110}.

\bibitem{Fily12}
\Name{Fily Y. \and Marchetti M.~C.} \REVIEW{Phys. Rev.
  Lett.}{108}{2012}{235702}.

\bibitem{Redner13}
\Name{Redner G.~S., Hagan M.~F. \and Baskaran A.} \REVIEW{Phys. Rev.
  Lett.}{110}{2013}{055701}.

\bibitem{Palacci13}
\Name{Palacci J., Sacanna S., Steinberg A.~P., Pine D.~J. \and Chaikin P.~M.}
  \REVIEW{Science}{339}{2013}{936}.

\bibitem{Buttinoni13}
\Name{Buttinoni I., Bialk\'e J., K\"ummel F., L\"owen H., Bechinger C. \and
  Speck T.} \REVIEW{Phys. Rev. Lett.}{110}{2013}{238301}.

\bibitem{Reichhardt15}
\Name{Reichhardt C. \and Reichhardt C. J.~O.} \REVIEW{Phys. Rev.
  E}{91}{2015}{032313}.

\bibitem{Cates15}
\Name{Cates M.~E. \and Tailleur J.} \REVIEW{Annual Review of Condensed Matter
  Physics}{6}{2015}{219}.

\bibitem{Tailleur09}
\Name{Tailleur J. \and Cates M.~E.} \REVIEW{EPL}{86}{2009}{60002}.

\bibitem{Fily15}
\Name{Fily Y., Baskaran A. \and Hagan M.~F.} \REVIEW{Phys. Rev.
  E}{91}{2015}{012125}.

\bibitem{Sepulveda17}
\Name{Sep\'ulveda N. \and Soto R.} \REVIEW{Phys. Rev.
  Lett.}{119}{2017}{078001}.

\bibitem{Solon15}
\Name{Solon A.~P., Fily Y., Baskaran A., Cates M.~E., Kafri Y., Kardar M. \and
  Tailleur J.} \REVIEW{Nature Phys.}{11}{2015}{673}.

\bibitem{Ray14}
\Name{Ray D., Reichhardt C. \and Reichhardt C. J.~O.} \REVIEW{Phys. Rev.
  E}{90}{2014}{013019}.

\bibitem{Ni15}
\Name{Ni R., Cohen~Stuart M.~A. \and Bolhuis P.~G.} \REVIEW{Phys. Rev.
  Lett.}{114}{2015}{018302}.

\bibitem{Kjeldbjerg21}
\Name{Kjeldbjerg C.~M. \and Brady J.~F.} \REVIEW{Soft Matter}{17}{2021}{523}.

\bibitem{Galajda08}
\Name{Galajda P., Keymer J., Dalland J., Park S., Kou S. \and Austin R.}
  \REVIEW{J. Modern Optics}{55}{2008}{3413}.

\bibitem{Reichhardt17a}
\Name{Reichhardt C. J.~O. \and Reichhardt C.} \REVIEW{Ann. Rev. Condens. Matter
  Phys.}{8}{2017}{51}.

\bibitem{Reichhardt14}
\Name{Reichhardt C. \and Olson~Reichhardt C.~J.} \REVIEW{Phys. Rev.
  E}{90}{2014}{012701}.

\bibitem{Chepizhko15}
\Name{Chepizhko O. \and Peruani F.} \REVIEW{Eur. Phys. J. Spec.
  Top.}{224}{2015}{1287}.

\bibitem{Morin17}
\Name{Morin A., Desreumaux N., Caussin J.-B. \and Bartolo D.} \REVIEW{Nature
  Phys.}{13}{2017}{63}.

\bibitem{Bhattacharjee19a}
\Name{Bhattacharjee T. \and Dutta S.~S.} \REVIEW{Nature
  Commun.}{10}{2019}{2075}.

\bibitem{Reichhardt17}
\Name{Reichhardt C. \and Reichhardt C. J.~O.} \REVIEW{Rep. Prog.
  Phys.}{80}{2017}{026501}.

\bibitem{Benassi11}
\Name{Benassi A., Vanossi A. \and Tosatti E.} \REVIEW{Nature
  Commun.}{2}{2011}{236}.

\bibitem{Buchler03}
\Name{B\"uchler H.~P., Blatter G. \and Zwerger W.} \REVIEW{Phys. Rev.
  Lett.}{90}{2003}{130401}.

\bibitem{Lewenstein07}
\Name{Lewenstein M., Sanpera A., Ahufinger V., Damski B., Sen(De) A. \and Sen
  U.} \REVIEW{Adv. Phys.}{56}{2007}{243}.

\bibitem{Gross17}
\Name{Gross C. \and Bloch I.} \REVIEW{Science}{357}{2017}{995}.

\bibitem{Brunner02}
\Name{Brunner M. \and Bechinger C.} \REVIEW{Phys. Rev.
  Lett.}{88}{2002}{248302}.

\bibitem{Brazda18}
\Name{Brazda T., Silva A., Manini N., Vanossi A., Guerra R., Tosatti E. \and
  Bechinger C.} \REVIEW{Phys. Rev. X}{8}{2018}{011050}.

\bibitem{Coppersmith82}
\Name{Coppersmith S.~N., Fisher D.~S., Halperin B.~I., Lee P.~A. \and Brinkman
  W.~F.} \REVIEW{Phys. Rev. B}{25}{1982}{349}.

\bibitem{Bak82}
\Name{Bak P.} \REVIEW{Rep. Prog. Phys.}{45}{1982}{587}.

\bibitem{Harada96}
\Name{Harada K., Kamimura O., Kasai H., Matsuda T., Tonomura A. \and
  Moshchalkov V.~V.} \REVIEW{Science}{274}{1996}{1167}.

\bibitem{Reichhardt98a}
\Name{Reichhardt C., Olson C.~J. \and Nori F.} \REVIEW{Phys. Rev.
  B}{57}{1998}{7937}.

\bibitem{OrtizAmbriz19}
\Name{Ortiz-Ambriz A., Nisoli C., Reichhardt C., Reichhardt C. J.~O. \and
  Tierno P.} \REVIEW{Rev. Mod. Phys.}{91}{2019}{041003}.

\bibitem{Bohlein12}
\Name{Bohlein T., Mikhael J. \and Bechinger C.} \REVIEW{Nature
  Mater.}{11}{2012}{126}.

\bibitem{Vanossi13}
\Name{Vanossi A., Manini N., Urbakh M., Zapperi S. \and Tosatti E.}
  \REVIEW{Rev. Mod. Phys.}{85}{2013}{529}.

\bibitem{Reichhardt20}
\Name{Reichhardt C. \and Reichhardt C. J.~O.} \REVIEW{Phys. Rev.
  E}{102}{2020}{042616}.

\bibitem{Reichhardt21}
\Name{Reichhardt C. \and Reichhardt C. J.~O.} \REVIEW{Phys. Rev.
  E}{103}{2021}{022602}.

\bibitem{Lowen16}
\Name{L{\" o}wen H.} \REVIEW{Eur. Phys. J. Spec. Top.}{225}{2016}{2319}.

\bibitem{Han17}
\Name{Han M., Yan J., Granick S. \and Luijten E.} \REVIEW{Proc. Natl. Acad.
  Sci. (USA)}{114}{2017}{7513}.

\bibitem{Reichhardt19}
\Name{Reichhardt C. \and Reichhardt C. J.~O.} \REVIEW{J. Chem.
  Phys.}{150}{2019}{064905}.

\bibitem{Shapere12}
\Name{Shapere A. \and Wilczek F.} \REVIEW{Phys. Rev. Lett.}{109}{2012}{160402}.

\bibitem{Yao20}
\Name{Yao N.~Y., Nayak C., Balents L. \and Zaletel M.~P.} \REVIEW{Nature
  Phys.}{16}{2020}{438}.

\bibitem{Libal20}
\Name{Lib\'al A., Bal\'azs T., Reichhardt C. \and Reichhardt C. J.~O.}
  \REVIEW{Phys. Rev. Lett.}{124}{2020}{208004}.

\bibitem{Bhushan13}
\Name{Bhushan B.} \Book{Introduction to Tribology, 2nd Edition} (John Wiley \&
  Sons, West Sussex, United Kingdom) 2013.

\bibitem{Derenyi95}
\Name{Der\'enyi I. \and Vicsek T.} \REVIEW{Phys. Rev. Lett.}{75}{1995}{374}.

\bibitem{deSouzaSilva06}
\Name{de~Souza~Silva C.~C., Van~de Vondel J., Zhu B.~Y., Morelle M. \and
  Moshchalkov V.~V.} \REVIEW{Phys. Rev. B}{73}{2006}{014507}.

\bibitem{Lu07}
\Name{Lu Q., Reichhardt C. J.~O. \and Reichhardt C.} \REVIEW{Phys. Rev.
  B}{75}{2007}{054502}.

\bibitem{Potiguar14}
\Name{Potiguar F.~Q., Farias G.~A. \and Ferreira W.~P.} \REVIEW{Phys. Rev.
  E}{90}{2014}{012307}.

\bibitem{Lozano16}
\Name{Lozano C., ten Hagen B., L{\" o}wen H. \and Bechinger C.} \REVIEW{Nature
  Commun.}{7}{2016}{12828}.

\bibitem{McDermott16}
\Name{McDermott D., Olson~Reichhardt C.~J. \and Reichhardt C.} \REVIEW{Soft
  Matter}{12}{2016}{8606}.

\bibitem{Reichhardt13a}
\Name{Reichhardt C. \and Reichhardt C. J.~O.} \REVIEW{Phys. Rev.
  E}{88}{2013}{042306}.

\bibitem{Reichhardt20b}
\Name{Reichhardt C. \and Reichhardt C. J.~O.} \REVIEW{Phys. Rev.
  E}{101}{2020}{062602}.

\bibitem{Vicsek95}
\Name{Vicsek T., Czir\'ok A., Ben-Jacob E., Cohen I. \and Shochet O.}
  \REVIEW{Phys. Rev. Lett.}{75}{1995}{1226}.

\bibitem{Drocco12}
\Name{Drocco J.~A., Olson~Reichhardt C.~J. \and Reichhardt C.} \REVIEW{Phys.
  Rev. E}{85}{2012}{056102}.

\end{thebibliography}

\end{document}